# Mobile GIS and Open Source Platform Based on Android: Technology for System Pregnant Women

Ayad Ghany Ismaeel, Nur Gaylan Hamead

**Abstract**— the statistic of World Health Organization shows at one year about 287000 women died most of them during and following pregnancy and childbirth in Africa and south Asia. This paper suggests system for serving pregnant women using open source based on Android technology, the proposed system works based on mobile GIS to select closest care centre or hospital maternity on Google map for the pregnant woman, which completed an online registration by sending SMS via GPRS network (or internet) contains her name and phone number and region (Longitude and Latitude) and other required information the server will save the information in server database then find the closest care centre and call her for first review at the selected care centre, the  proposed system allowed the pregnant women from her location (home, market, etc) can send a help request in emergency cases (via SMS by click one button) contains the ID for this pregnant woman, and her coordinates (Longitude and Latitude) via GPRS network, then the server will locate the pregnant on Google map and retrieve the pregnant information from the database. This information will be used by the server to send succoring to pregnant woman at her location and at the same time notify the nearest hospital and moreover, the server will send SMS over IP to inform her husband and the hospital doctors. Implement and applied this proposed system of pregnant women shows more effective cost than other systems because it works in economic mode (SMS), and the services of proposed system are flexible (open source platform) as well as rapidly (mobile GIS based on Android) achieved e.g. locally registration, succoring in emergency cases, change the review date of pregnant woman, addition to different types of advising according to pregnancy.

————————————  ◆  ————————————

## 1 Introduction

World Health Organization says about 800 women die from complications of pregnancy or childbirth-related worldwide at day. 99% of maternal deaths in poor countries and only 46% of their women benefit from skilled care during childbirth. Solving this problem began with increasing of using mobile telephone, these advantage subscribers are taken in the country to improve the health of their citizens and overcome existing communication barriers by broadcasting SMS text messages to all mobile telephone for the pregnant women via pregnancy care advice [1].

Realy there is big problem needed to solve by suggest system for medical reasons using to cover pregnant women requirements start with registration in closest care center, advice, alarm when there is help request from pregnant woman by displayed on the map and succoring health system can track the pregnant only when she need succoring the system using open source technology that using smart phones based on Android. IPhone, iPad platform: It is inexpensive and easy to develop for, it is available to millions of potential users worldwide and it has fewer limitations than other platforms [2]. Here needed to explain the technology which is employed for expacted enivernoment as follow:

———————————————————
- *Ayad Ghany Ismaeel, PhD in computer science. currently is professor at department of Information System Engineering, Erbil Technical Engineerin College- Hawler Polytechnic University (previous FTE- Erbil), Iraq. PH-009647703580299. E-mail:* dr_a_gh_i@yahoo.com, *Alternative E-mail:* dr.ayad.ghany.ismaeel@gmail.com
- *Nur Gaylan Hamead is MSc Student in Computer Science, College of Science, Salahaddin University.Email:* yosifasn@yahoo.com

## 1.1 Android Technology

Android is an open source software toolkit for mobile phones that was Created by Google and the Open Handset Alliance. It's inside millions of cell phones and other mobile devices, making Android a major platform for application developers. Whether you're a hobbyist or a professional programmer [3], open source software is currently one of the most debated phenomena in the Software industry, both theoretically and empirically. At the most basic level, the term open source software simply means software for which the source code is open and available.

Smart phones and tablets become more popular; the operating systems for those devices become more important. Android is such an operating system for low powered devices, that on battery and are full of hardware like Global Positioning System (GPS) receivers, cameras, light and orientation sensors, WiFi and UMTS (3G telephony) connectivity and a touch screen. Like all operating systems, Android enables applications to make use of the hardware features through abstraction and provide a defined environment for applications [4]. Android was sold to Google in 2005; it is based on a modified Linux 2.6 kernel. Google, as well as other members of the Open Handset Alliance (OHA) collaborated on Android (design, development, distribution). Currently, the Android Open Source Project (AOSP) is governing the Android maintenance and development cycle [5].

Google wanted; Android to be open and free; hence, most of the Android code was released under the open source Apache License, which means that anyone who wants to use Android can do so by downloading the full Android source code. Moreover, vendors (typically hardware manufacturers) can add their own proprietary extensions to Android and customize Android to differentiate their products from others. This simple development model makes Android very attractive and has thus piqued the interest of many vendors. To summarize, the Android operating environment can be labeled as [5]:

1.  An open platform for mobile development
2.  A hardware reference design for mobile devices

3. A system powered by a modified Linux 2.6 kernel
4. A run time environment
5. An application and user interface (UI) framework [6].

## 1.2 Android System Architecture

The Android software stack shown in Fig. 1, it can be subdivided into five layers: The kernel and low level tools, native libraries, the Android Runtime, the framework layer and on top of all the applications [5].

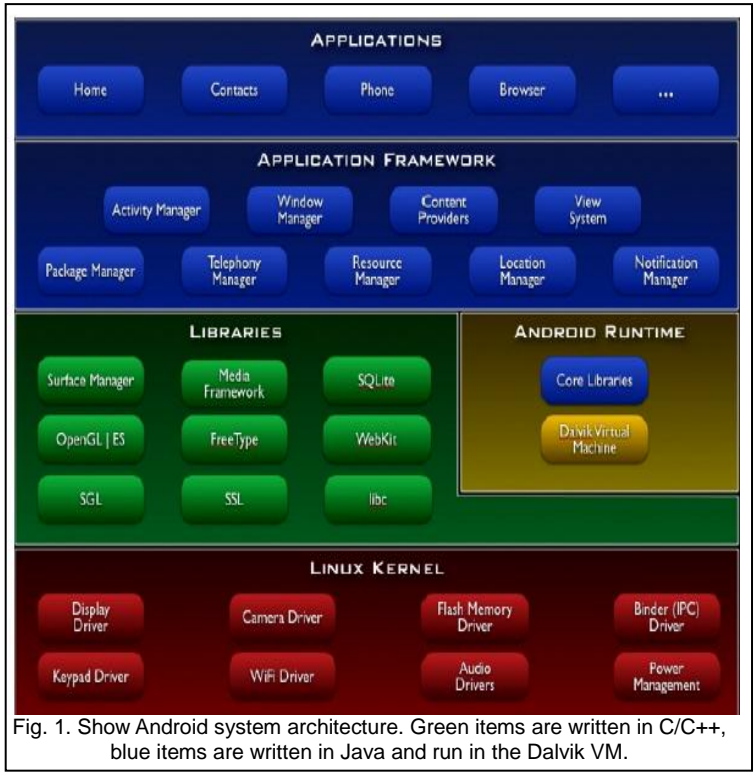

Fig. 1. Show Android system architecture. Green items are written in C/C++, blue items are written in Java and run in the Dalvik VM.

## 1.3 Open Source Platform

Bruce Perens defines that Open Source is a Specification of what is permissible in a software license for that software to be referred to as Open Source. Developing done by "Any one who contributes to the open Source project is an open source developer." such as a User of the software, a developer who develops the Software, a debugger or hobbyist who likes spending time on open source, or a promoter who funds such a Development. Eric states that developers are attracted towards open source development because that gives them an opportunity to demonstrate their ability. So they voluntarily select a project and start contributing. When programmer's, code gets accepted, it boosts their ego and they get recognized for their effort in the community, Peer recognition creates reputation and a reputation as a good programmer is a great achievement [6].

Open source developers are involved in a variety of activities such as designing, coding, debugging and utilizing. Each activity occurs simultaneously. Parallel development and debugging is the key to open source success, users also play a vital role in the debugging Process by reporting bugs to developers or sometimes fixing it themselves. Developers are well aware that users are the best beta testers [6].

## 2 RELATED WORK

Ayad Ghany Ismaeel and Emad Khadhm Jabar [ 2013] suggested mHealth system for serving pregnant women, that proposed system is first an effective mHealth system works base on mobile GIS to select adjacent care centre or hospital maternity on Google map at online registration for woman pregnant, that is done when the pregnant woman will send SMS via GPRS network contains her ID and coordinates (Longitude and Latitude) the server when receive it will search database support that system and using the same infrastructure for help the pregnant women at her location (home, market, etc) in emergency cases when the woman send SMS contains her coordinates for succoring. Implement the proposed pregnant women system shows more effective from view of cost than other systems because it works in economic (SMS) mode and from view of serve the system can easy and rapidly manage when achieving locally registration, succoring in emergency cases, change the review date of pregnant woman, as well as different types of advising [7].

Ayad Ghany Ismaeel and Sanaa Enwaya Rizqo [2013] offered a succoring system controlled by the patient (e.g. pregnant women, child, young, etc) based on the patients' location. The proposed system is the first tracking system using mobile GIS based on WCF technology to offer online succoring (24 hour a day), but really works only when the patient sends request for succoring. The patients will send a request (SMS by click one button) contains his ID, Longitude and Latitude via GPRS network to a web server containing a database, which the patient was registered previously on it. Then the server will locate the patient on Google map and retrieve the patient's information from the database. This information will be used by the server to send succoring facility and notify the nearest and most suitable ESC; moreover, the server will send SMS over IP to inform the patient emergency contacts and emergency hospital. The optimal productivity for proposed succoring system appears in handling a large number of requests within short period at rate of one request/need succoring per sec as result of using mobile GIS based on WCF technology. Furthermore, the process of request and reply for emergency cases of the patients achieved in cost-effective way due to this technology, which allow sending data (SMS over IP) via Internet using GPRS network. The proposed system can be implemented in a minimum configuration (hardware and software) to minimize the overall cost of operation and manufacturing [8].

Bangladesh experience [2010] allows the mobile users at 2010 subscribing, at a reduced rate, to S SMS service that broadcasts messages in health topics. Health workers in communities throughout the country can advise Bangladesh as case study via SMS the patients through their mobile telephones. From those patients the pregnant woman can register their mobile numbers to receive prenatal advice [1].

The whole previous eHealth/mHealth systems for pregnant women are offers advising using mobile or mobile GIS in their services but each of them has weaknesses like there isn't determine closest care (health) center for pregnant woman from her phone, call for review and changing the date of next review using mobile, advising relate to each trimester of pregnancy, there isn't another option the system can serve the pregnant women via Internet (ISP), etc. The motivation overcome on the problems above and reaching to system for serving pregnant women can achieve easy registration (from their homes) can select the nearest care center for her and can obtain succoring in emergency case at her location by succoring facility (car, helicopter, boat life) take her to nearest hospital.

For the suggested health system (pregnanet women) must be based on the following techniques and modes [7]:

**A. Mobile GIS:** is the expansion of GIS technology from the office into the field. A mobile GIS enables field-based personnel to capture, store, and update, manipulate, analyze, and display geographic information. It is an integrated software/hardware framework for the access of geospatial data and services through mobile devices via wireless networks. Mobile GIS integrates one or more of the following technologies (mobile devices, Global Position System (GPS) and wireless communications for Internet GIS access) as shown in Fig. 2 [8].

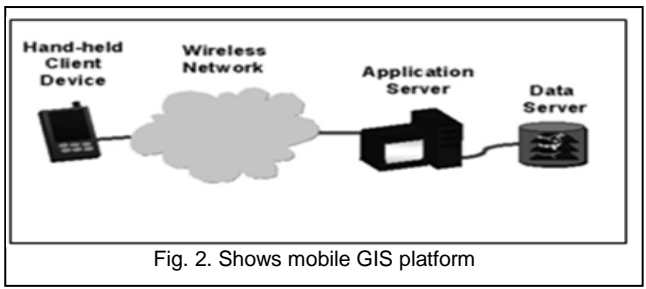

Fig. 2. Shows mobile GIS platform

The proposed system will base on supporting of a mobile build-in GPS and A- GPS technique on device Samsung Galaxy S III (OS-Upgrades to 4.1.2 "Jelly Bean" (July 2013) Touch Wiz "Nature UX" GUI.

**B. Modes of transmission as follow [8]:**

**i.** GPRS: the mobile terminal sends the External data from GPS satellites through GPS data channel to a special TCP/IP server (a PC with dynamic IP address) linked to the internet.

**ii.** SMS over IP (SoIP) mode: The Short Message Service (SMS) is one of the most successful services in existing cellular networks. The SMS provides a means of sending messages of limited size to and from Global System for Mobile Communications (GSM) or Universal Mobile Telecommunications System (UMTS) phones as shows in Fig 3.

## 3 PROPOSED SYSTEM OF PREGNANT WOMEN

The architecture of suggested health system for succoring the pregnant women involves multiple stages one of them Emergency Manage Center EMC, Fig. 4 shows the architecture of this proposed system:

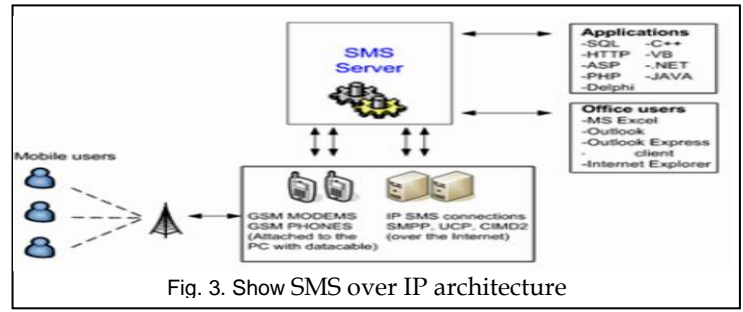

Fig. 3. Show SMS over IP architecture

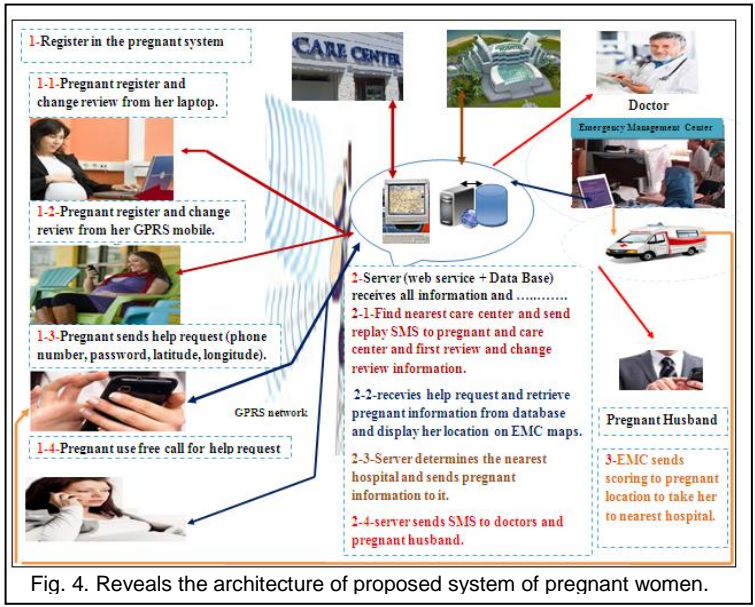

Fig. 4. Reveals the architecture of proposed system of pregnant women.

The general tasks of pregnant women system can summarized as flowchart shown in Fig.5.

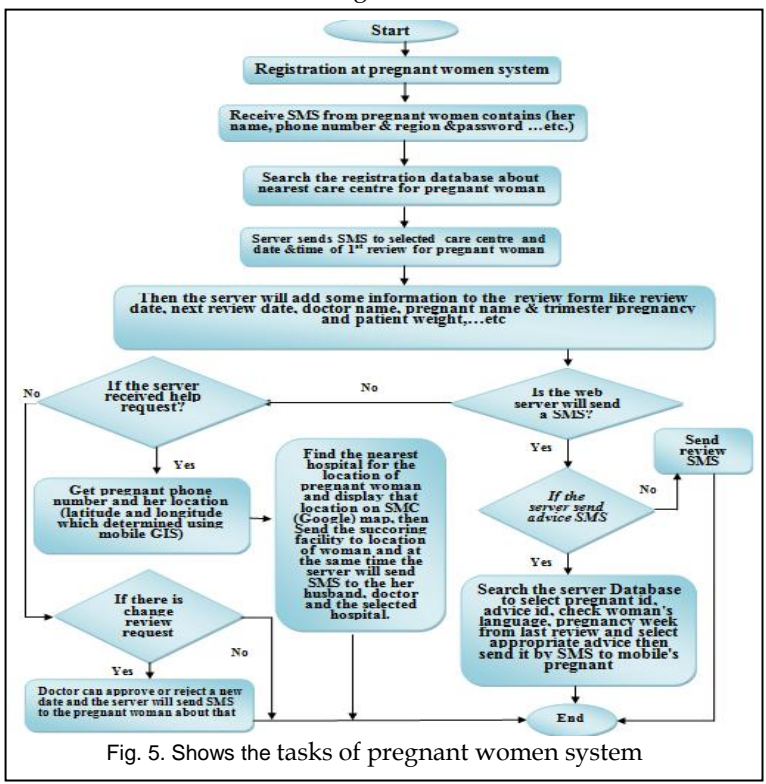

Fig. 5. Shows the tasks of pregnant women system

Fig. 6 reveals one of the important services in this proposed system, which is request help, shown in the following flowchart.

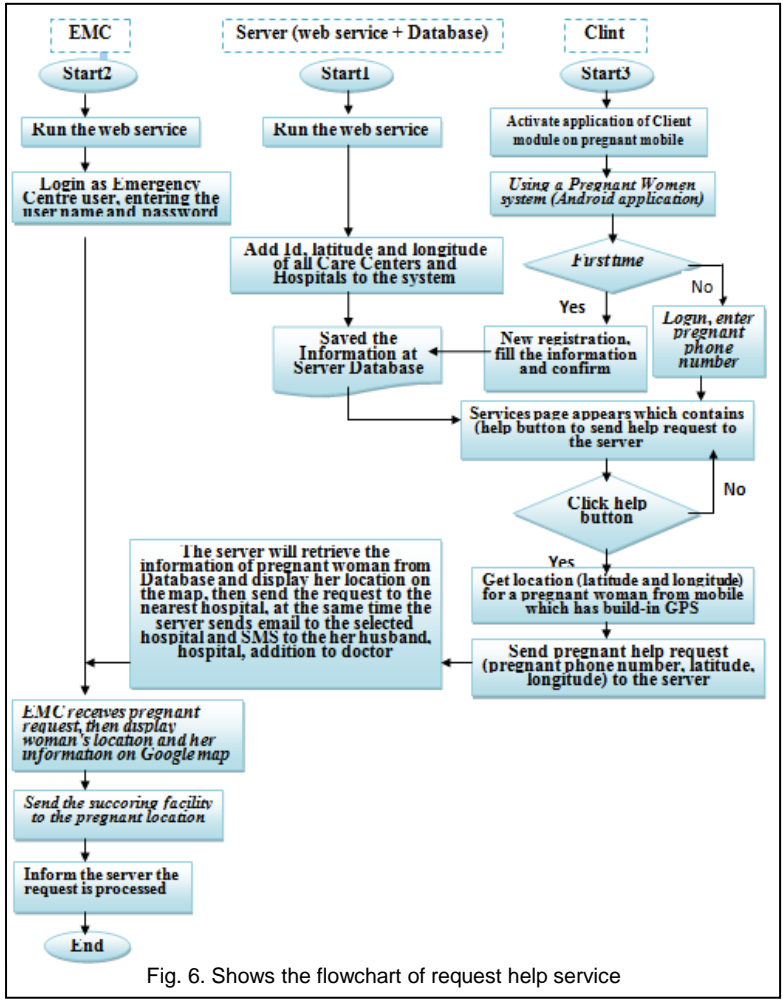

Fig. 6. Shows the flowchart of request help service

The proposed system of serving the pregnant women using mobile GIS based on open source platform contains five parts are:

**A. The Client Side:** In this system the client is a mobile hold by the pregnant women. e.g. Samsung Galaxy S III a mobile buildin GPS receiver on device (OS-Upgrades to 4.1.2 "Jelly Bean" (July 2013) Touch Wiz "Nature UX" GUI) and GPRS transmitter work over GSM network is used in this system because it is more friendly with open source platform and support A- GPS technique, this compatibility will avoid the conflict in dealing with the software. The mobile application was developed with java language in Android studio (I/O preview) 0.3.2 built on 2013 by using java SDK7 and android SDK. Fig. 7 shows a flowchart of the main steps of running the client/mobile application under pregnant women system.

**B. The Server Side:** is a PC provided with an internet connection (Internet Service Provider ISP) and has dynamic IP address, supported by Windows operating system, Apache2.2 server, php5 and MySQL database, this server will use for saving the incoming data from mobile. The server should be running 24 hours a day, to provide different services to the pregnant woman (register, change a date of review, first review, weekly advice and offer succoring in emergency cases). The server works automatically when receiving the request from the client (mobile of pregnant woman) and sends it to the nearest care center in registration case and sends it to the nearest hospital in request help case, and then send help request to the EMC also sends the SMS to the doctors of the hospital addition to husband of pregnant woman and emergency hospital. The main tasks of the web service for this side explained in Fig. 8.

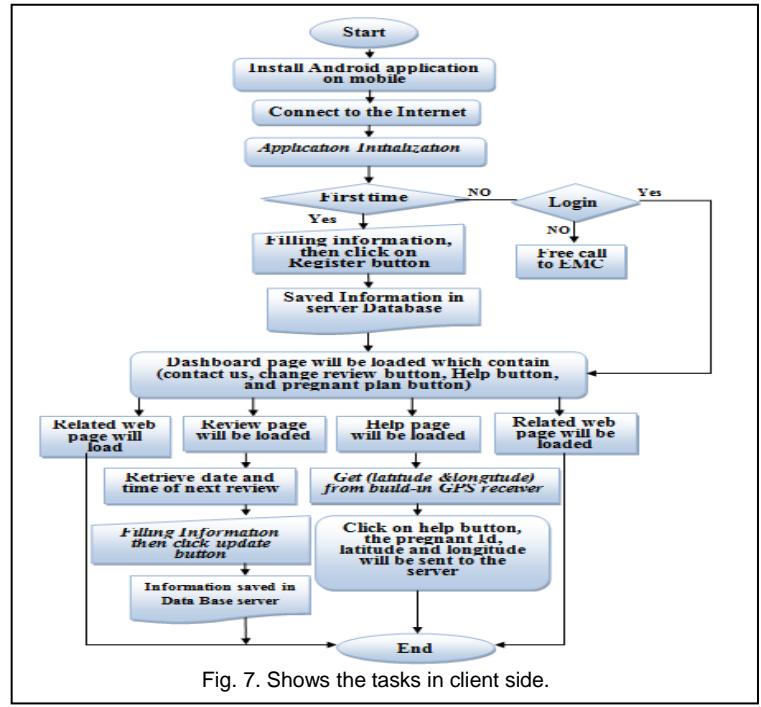

Fig. 7. Shows the tasks in client side.

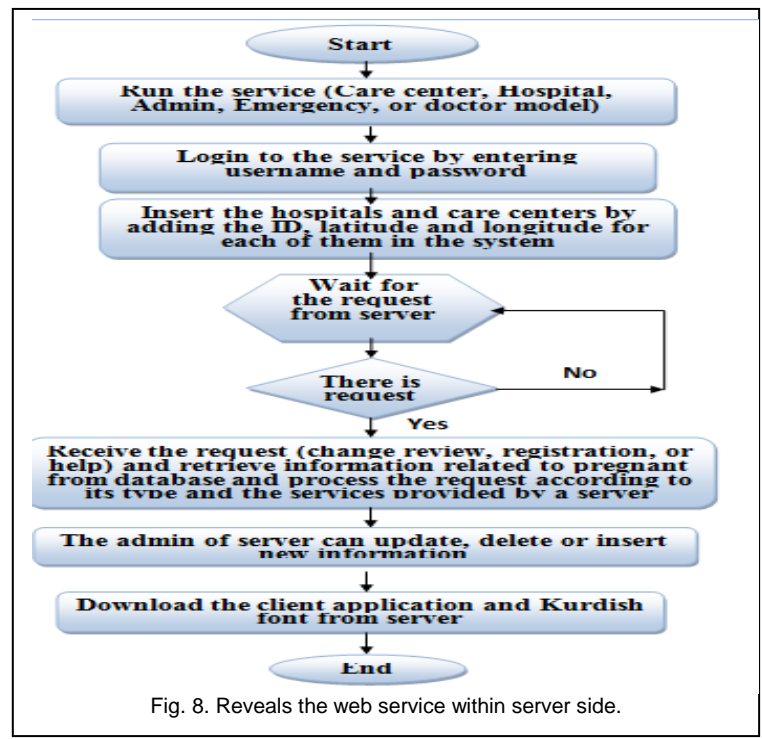

Fig. 8. Reveals the web service within server side.

Fig. 9 shows the (10) tables within server database (called registration), which created for the proposed system of preg-nant women, and the relation between these tables.

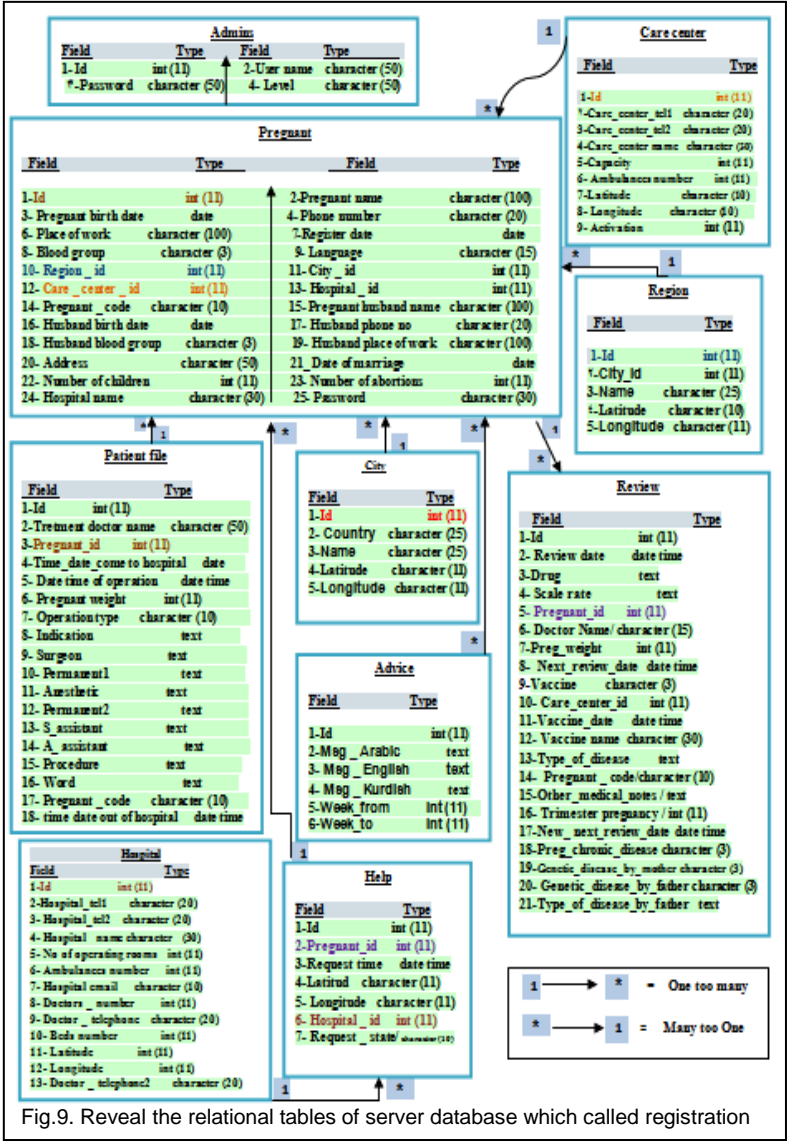

Fig.9. Reveal the relational tables of server database which called registration

**C. EMC Center:** this center created to track the pregnant woman when she send request help at emergancy cases, the system will show her location to EMC on Google map and then the EMC sends team of succoring to transfer her to the nearest hospital.

**D. Hospital and its' doctor:** the hospital receives SMS of emergency request from the server and informs the doctor with more information about the pregnant which was sent that request, and the doctor is filled the required information about the pregnant that transferred to the hospital at file patient, after inserting username and password of the doctor. The main tasks of this part shown in Fig. 10.

**E. E-Care Center:** the care center achieves scheduling to the review time for pregnant woman and applies weekly health care and advice for each pregnant woman according the trimester of pregnancy. Fig. 11 reveals the tasks under this part.

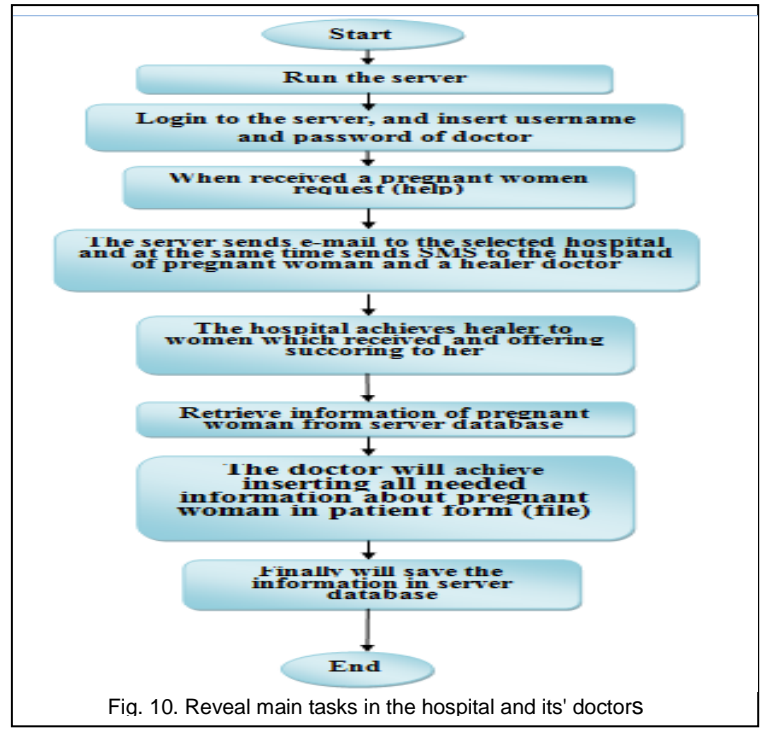

Fig. 10. Reveal main tasks in the hospital and its' doctors

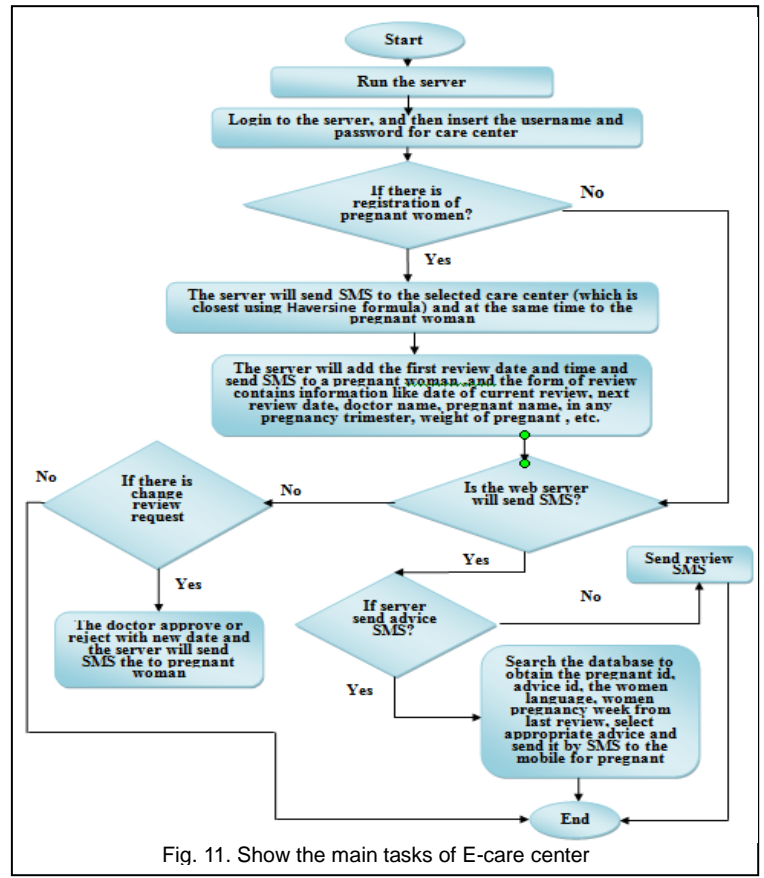

Fig. 11. Show the main tasks of E-care center

## 4 EXPERIMENTAL RESULTS

The web service in this system called (pregnant service) which was developed with php language in Dreamweaver Version 9.0_2007 based on open source platform.

## 4.1 Technologies Required

There are several important technologies that must be available to implement this proposed system as follow, these technologies will use to cover the platform of the pregnant women system which is shown in Fig. 12:

1) MySQL: For creating server database (registration) of with (10) relational tables as referring in section 3; B.
2) Open source platform.
3) Google Maps API V3.1.
4) Java language.
5) PHP 5 language.
6) Android studio for android phone.
7) Java SDK7.
8) Android SDK.
9) Genymotion for android vm.
10) Dreamweaver Version 9.0_2007 for php language.

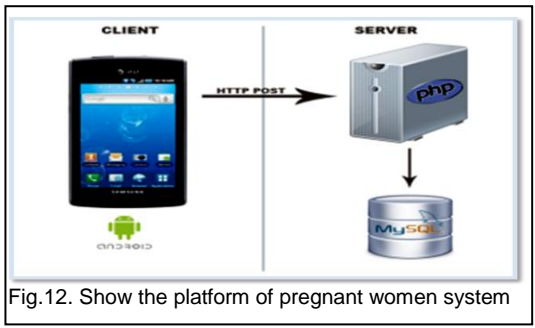

Fig.12. Show the platform of pregnant women system

## 4.2 Web Service Android Applications as Client Side

The woman who request system's services; can ask the admin directly or by SMS, and then the adminstrator will send the programs of client side (Android application) to the pregnant woman, which can install it on her smart mobile; after that the woman can do registration to obtain the services (applications) which available on this system. Fig. 13 reveals these web service (login for registration, call for help, change review, contact with admin, etc) as shown the system's GUI achieves in English additional to Kurdish and Arabic locally languages.

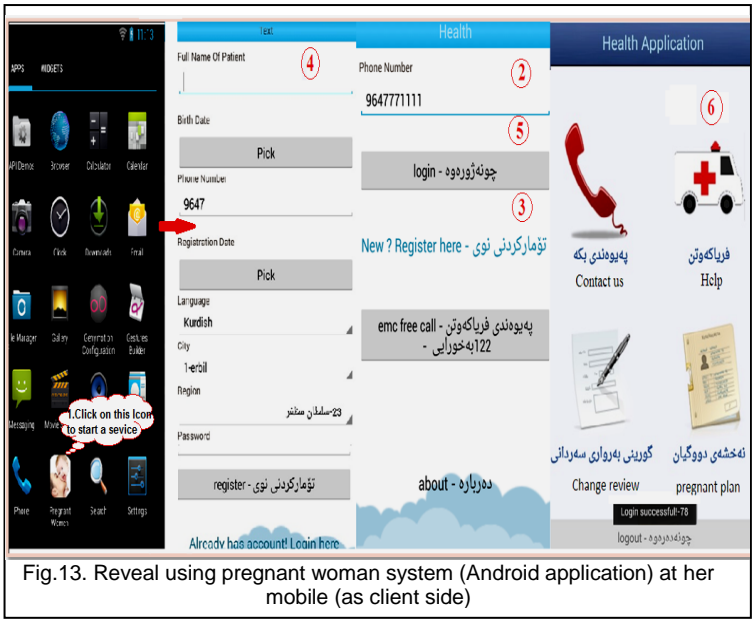

Fig.13. Reveal using pregnant woman system (Android application) at her mobile (as client side)

Another important service can seen in this side, the woman can call for help (in emergency case) from her location (in/out house in market, job, etc) to implement that the woman for easy must be click only one button, If the demand of the pregnant woman is done will see a message as shown in Fig. 14; A, otherwise will see message as shown in Fig. 14; B.

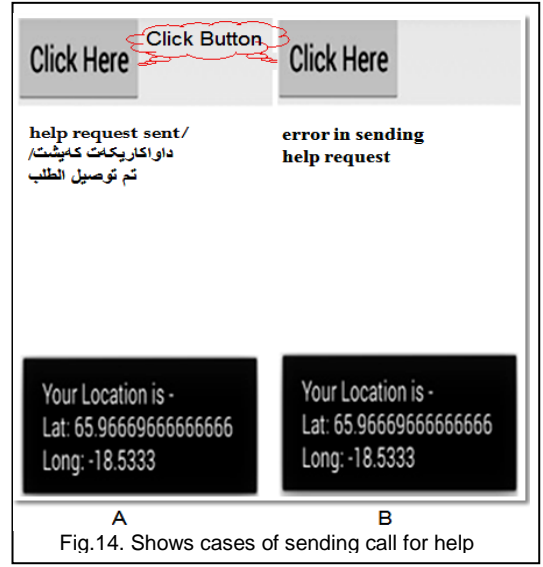

Fig.14. Shows cases of sending call for help

## 4.3 Web Service at Server Side

The server has multiple tasks one of it searching the server database, e.g. to obtain statistical reports, provide advice or find closest care centre, etc. When the server receives a registration request the system will use Haversine formula (which is an equation important in navigation, giving great-circle distances between two points on a sphere from their longitudes and latitudes) to find the nearest hospital or care center and will inform the selected care center and the pregnant woman by SMS for first review. The pseudo code of this formula in php shows as follow:

```
Class Haversine
{   Private $radian;
    Private $sphere Radius;
    Private $startLatDeg;
    Private $startLongDeg;
    Private $endLatDeg;
    Private $endLongDeg;
Public $distance;
    Public function __construct ()
    {   $this->radian = M_PI / 180;
        $this->sphere Radius = 6372.797;
$this->calculate Distance ();
        Return $this->distance;   } }
Public function get Distance ($startLat, $startLong, $endLat, $endLong)
{   $this->setStartLat ($startLat);
        $this->setStartLong ($startLong);
        $this->setEndLat ($endLat);
        $this->setEndLong ($endLong);   }
```

As refer before the proposed system offers service in server side, which is an advising to the pregnant women, the diagram for this service shown in Fig. 15.

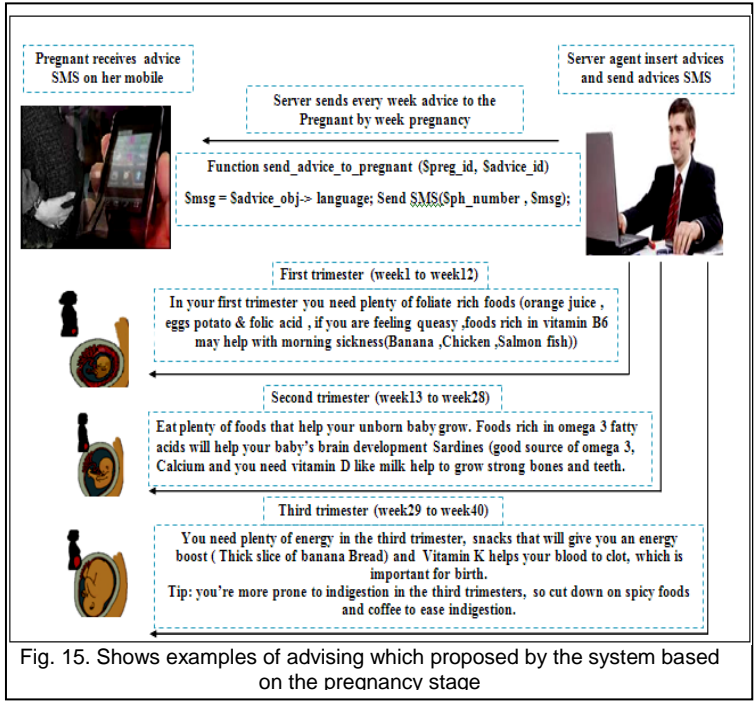

Fig. 15. Shows examples of advising which proposed by the system based on the pregnancy stage

The server's service start (in an emergency case) when recieved SMS from the pregnant woman, and then the server will retrieve her information and determine her location at Google map of the EMC, and then EMC will send succoring (car, boat-life, helicopter, etc) to this woman at her location as shown in Fig. 16, and in the same time the server will send SMS to the hospital, doctor of this pregnant woman, her husband, etc.

Fig. 16. Shows help request from pregnant woman on the Google map.

### 4.4 Discussion the Results

Table 1 shows comparing the results of proposed system for pregnant women with other system related.

TABLE 1
REVEALS COMPARISON OF PROPOSED SYSTEM WITH OTHER RELATED SYSTEMS

| Proposed system of pregnant women | Ayad Gh. Ismaeel & Eam Kh. Jabar [7] | Ayad Gh. Ismaeel & Sanaa E. Rizqo [8] | Bangladesh experience [1] |
|---|---|---|---|
| Mobile GIS and open source platform based on Android Technology | Mobile GIS based on Windows platform | Mobile GIS based on Windows platform & WCF Technology | Not exist this service |
| Data transferring between client and server via GPRS and Internet | SMS via GPRS | Data transferring between client and server WCF technology (Request & reply) via Internet | SMS via GPRS but from server to pregnant women only |
| Registration and the request help of pregnant woman to the nearest Care Centre or hospital | Registration and send request to Emergancy Service Center ESC without caring to the distance | Registration and send patient request to nearest ESC to succor patient at shortest distance (time) | Only the registration exist |
| The system offers advice to pregnancy. | Exist this service | Not exist this service | Exist this service |
| Using smart mobile via web service | Using Windows mobile via web interface | Using Windows mobile via web service | Not exist this service |

## 5 CONCLUSIONS

The important conclusions which obtained from the proposed system for pregnant women as follow:

A. This pregnant women system is more effective than other systems because it supports locally registration for pregnant woman (from her home or place) and the server will find to her closest care centre, addition the system sends SMS for review, advising, interactive, etc via internet and GPRS network as economice mode.

B. The proposed system is very flexible and friendly in used because it based on open source platform and Android (smart) technology. Addition that allow mixing between technologies, e.g. using php scripts at server side, while using Java in client side. .

C. The system of pregnancy achieves higher productivity by selecting the nearest care centre to the pregnant women by search the server database and when the pregnant woman calls for help the system response at shortest time.

D. Using Andriod technology that support A-GPS, which stands for Assisted GPS helps a standalone

GPS unit to lock on to a satellite signal. Known as the TTFF (time-to-first-fix), this startup period can be challenging for GPS units where the satellite signal is weak or distorted by surrounding buildings.

E.   Using open source technology allows the users of Andriod (in client side) to update the application for the better.

## ACKNOWLEDGMENT

Thank to my wife (DR. NEMA SILAH ABDAL KAREEM) that helped me in the maturation of the idea of research and advices at the medical side, which has to do with their competence.